# Online measurement of optical fibre geometry during manufacturing

Maksim Shpak*[a], Sven Burger[b,c], Ville Byman[a], Kimmo Saastamoinen[d], Mertsi Haapalainen[e], Antti Lassila[a]

[a]MIKES Metrology, VTT Technical Research Centre of Finland Ltd, Espoo, Finland
[b]JCMwave GmbH, Berlin, Germany
[c]Zuse Institute Berlin, Berlin, Germany
[d]University of Eastern Finland, Joensuu, Finland
[e]Oplatek Group Oy, Leppävirta, Finland

## ABSTRACT

Online measurement of diameters and concentricities of optical fibre layers, and the coating layer in particular, is one of the challenges in fibre manufacturing. Currently available instruments can measure concentricity and diameter of layers offline, and are not suitable for precise monitoring or control of the manufacturing process in real time.

In this work, we use two laser beams, positioned orthogonally to illuminate the fibre from two sides, and calculate deviations from the expected geometry by analysing the scattering pattern. To measure the diffraction pattern we use two 8K linear array detectors, with the scattered light incident directly on the sensors. Each detector is capturing approximately 90° angular range directly behind the fibre. The two measurement channels are positioned at different heights.

The scattered pattern is modelled mathematically with finite-element and Fourier-modal methods, with various diameter and concentricity deviations. The sensitivities of the changes in the scattering pattern are identified in respect to these deviations. Since calculations are computationally intensive, the sensitivities are pre-calculated in advance, and the real-time measurement is based on pattern recognition. The symmetry of the pattern is used to differentiate between diameter and concentricity variations.

We performed online measurements with the prototype instrument in production conditions, and show that this method is sensitive enough to measure deviations of under 1 μm in diameter and concentricity of the coating layer.

## 1. INTRODUCTION

Optical fibre is manufactured by drawing it from a heated glass preform, which contains the layers of resulting fibre in appropriate diameter ratios. The geometry of the fibre is thus defined by the geometry of preform, and by various process parameters, such as feeding rate, drawing tension and temperature. One or more polymer coating layers are applied during the same process by guiding the fibre through a heated coating cup. One of the functions of these layers is to reduce the stress acted upon the glass layers when the fibre is bent. Deviations from the specified diameter of the coating layers or from their concentricity to the glass layers negatively affects the mechanical performance of the fibre. Online measurement of concentricity of fibre layers, and the coating layer in particular, is one of the challenges in fibre manufacturing.

Currently available instruments can measure concentricity and diameter of layers offline, typically by multi-directional laser scanning of the fibre immersed in index-matching oil, or by analysis of a polished fibre end, either directly with a microscope, or by using dark-field illumination. These methods are naturally not suitable for monitoring or controlling the manufacturing process in real time.

A fast online method has been long used to determine concentricity of the fibre and its coating, which is based on analysis of forward scattering of a laser beam that illuminates the fibre transversely[1]. In our work, we use the same measurement principle, and improve it with modern detection and data analysis to allow the measurements of wider variety of fibres and reach higher resolution.

In this paper, we present the results of an online measurement performed with a prototype instrument in a fibre production facility. We have modelled the fibre geometry with a finite-element method, and measured the same type of fibre in real-time, while causing various geometrical deviations in layers.

## 2. THEORY AND MODELLING

Typical optical fibre consists of a glass core, and one or more glass cladding layers with lower index of refraction. This structure forms a waveguide, and is determined by the geometry of the preform. The diameter of the fibre is adjusted during the drawing process by adjusting the tension and the feeding rate, and measured typically with a scanning laser type instrument. Additionally, the fibre is typically coated with one or more protective polymer layers, by guiding it through a coating cup. In calculations and experiments presented in this paper, the polymer coatings are transparent and have a higher index of refraction than the cladding layers. Figure 1 shows a typical layer structure of a fibre.

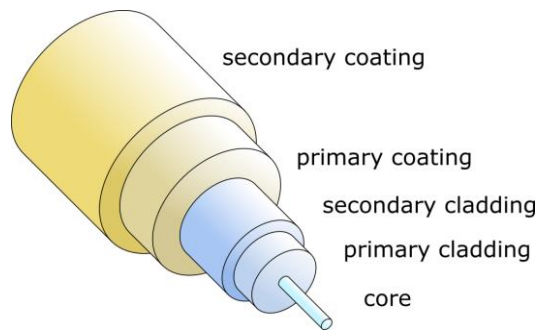

Figure 1. Layer structure of a typical optical fibre.

When the fibre is illuminated from the side, the light refracts as it propagates through the layers in the fibre. To the first approximation, the fibre acts as a lens with a very short focal length, and in the far field the light is dispersed into wide angles. Using a laser as the illumination source, the coherence length of which is much higher than the thicknesses of layers in the fibre, parts of the beam take different paths through the fibre, and interfere when they exit in different phases and angles. This produces dense patterns in the far field, as illustrated in figure 2.

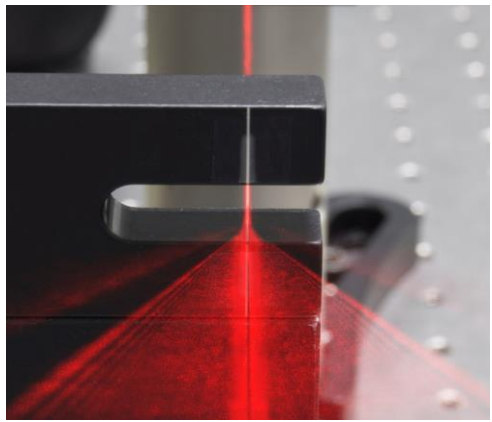

Figure 2. Interference pattern produced by illuminating the fibre with a laser beam (composite photo).

Modelling of the laser beam propagation has been performed with Fourier Modal[2,3] (FMM) and finite-element[4,5,6] methods (FEM). To validate the results of modelling, a comparison was done with the measured forward scattering profile, as reported previously[7]. A fibre with a 5µm core, 125µm cladding and 250µm coating layers was used. The comparison of intensity profiles as a function of angle is shown in figure 3. The rest of the models presented here were performed with the finite-element method.

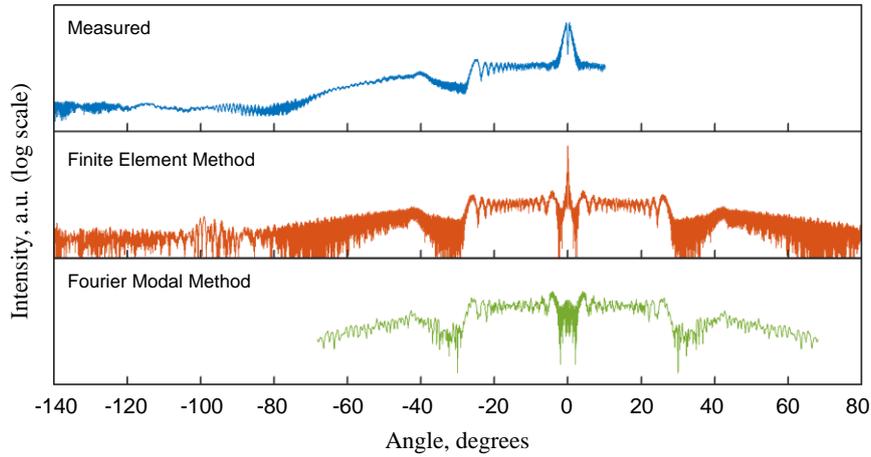

Figure 3. Comparison of modelling methods and measurements for one fibre type[4].

Positions of the oscillation peaks correlate with the geometrical dimensions in the fibre. E.g., an increase in the ratio of glass diameter to the coating diameter widens the intensity profile, and decrease makes it narrower. Similarly, deviations from concentricity of layers affect the symmetry of the intensity profile. We use these effects to extract the geometrical information.

Both the FFM and the FEM[5] are computationally intensive and not currently suitable for online-applications. A convergence study regarding a comparable setup involving a microstructured fibre and reporting typical reached numerical accuracies and computation times has been published previously[6]. The calculations in this implementation are done in advance, by modelling the scattering profile for a nominal fibre geometry and for every deviation of interest. From these results, the sensitivity parameters are calculated, which are used to determine the magnitude of deviations during the online measurements.

## 3. MEASUREMENT SETUP

A 3D model of the measurement setup is shown in figure 4. It consists of two horizontal detection channels, positioned orthogonally to each other. The measured fibre runs vertically through the middle of the instrument, passing through both measurement beams. As the laser beam is hitting the fibre, it produces a dense interference pattern in all directions around the fibre, and thus the detection channels are positioned at different heights to avoid cross talk. The light source is a 633 nm HeNe laser, coupled into a fibre splitter, and collimated with two microscope objectives.

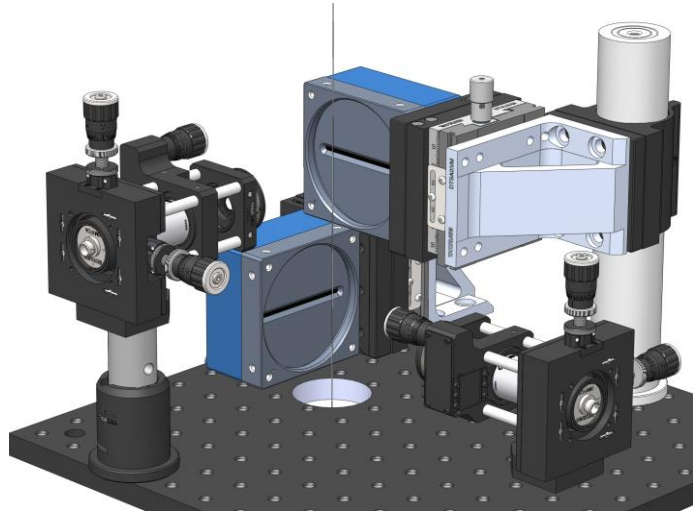

Figure 4. A model of the measurement setup.

To measure the diffraction pattern two 12-bit monochrome linear array detectors are used, positioned such that in the absence of the fibre the measurement beam of each channel is incident on the centre of the respective sensor. Each detector is capturing approximately 90° angular range directly behind the fibre. The resolution of each detector is 8192×1 pixels, with a pixel pitch of 7.04 µm. The achieved resolution in this configuration is between 0.015° (at the centres) to 0.008° (near the edges). Detectors were modified for the task by removing the protective glass in front of the sensors, because the thin glass was causing additional interference due to the coherence of the light source. With the sensors exposed, effective cleaning was problematic, and detectors were instead characterized for uniformity with a known illuminance source.

Data capturing is performed with a PC, through Gigabit Ethernet connection. A custom written software is controlling the cameras, collecting image data and performing analysis. The received raw data is first corrected with uniformity calibration, then a smoothing filtering is applied, and finally the positions of the centre of the fibre and of the two outermost intensity peaks are recorded, as illustrated in figure 5.

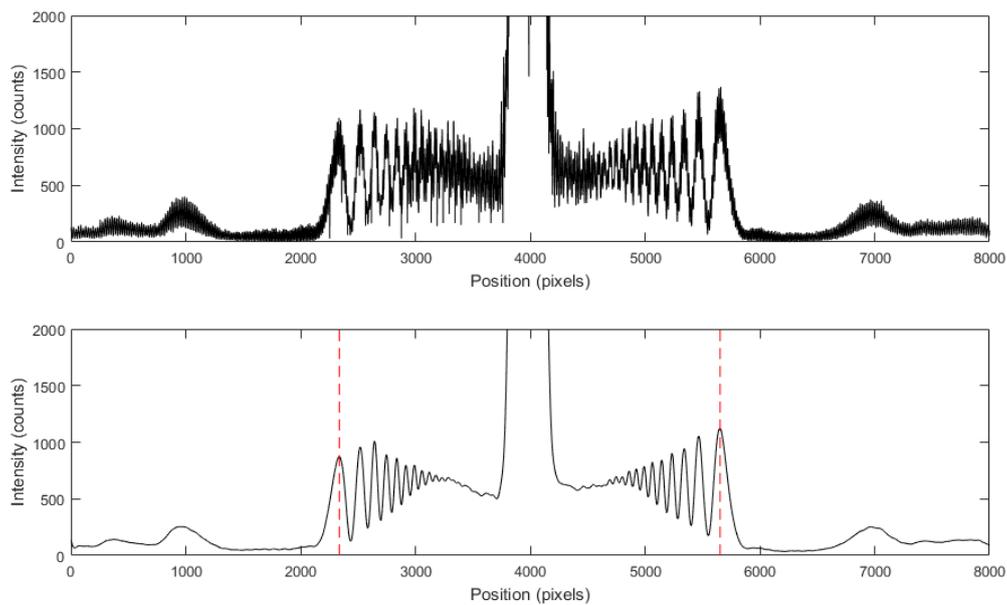

Figure 5. Example of collected camera data (above), filtered data (below), and detected peak positions (red dashed lines).

Test measurements were performed at an optical fibre manufacturing facility. The instrument was installed into a drawing tower, and the fibre was guided to through the instrument after the coating process. Diameters of the glass layer and the outer coating layer of the fibre were monitored with conventional instruments before and after the coating process. The varied parameters during the test were the thickness of the glass layer, thickness of the coating layer as well as concentricity of the glass inside the polymer coating. The measurement analysis uses the diameter of the glass layer as an input, and outputs the diameter and eccentricity of the outer coating layer. After the tests, the fibre samples were also measured with an offline instrument.

## 4. RESULTS

Online tests were performed for a single fibre type. The test fibre had a nominal glass diameter of 125 µm. Two polymer coating layers were applied, the inner coating diameter was 195 µm and the outer layer diameter was 260 µm. The expected positions of the interference peaks were derived for the nominal geometry with modelling, as well as their sensitivities to diameter changes and decentering. Sensitivity of the of the outermost interference peak position to the glass coating diameter change was 30.2 pixels/µm, to the coating diameter change 16.3 pixels/µm and to the eccentricity of the coating 67.1 pixels/µm.

In the first test, the effects of diameter variation were recorded. The diameter of outer coating layer was reduced by approximately 2 µm, while the diameter of the glass layer was kept constant. Next, the diameter of the glass layer was increased by 5 µm, and diameter of the coating was adjusted to be close to nominal. Table 1 shows the results of this test and comparison to values obtained with an offline instrument. Coating diameter values are highlighted with bold font, and in the second and third row, the values in brackets show the change compared to the first position. The ovality column is the absolute difference between diameters obtained from two measurement channels. The presented test measurement values are averages of several seconds of measurement.

Table 1. Diameter variation tests. All values are in µm.

|  | Test measurements | | | | Offline measurements | | | |
|---|---|---|---|---|---|---|---|---|
|  | **Coating diameter** | Eccentricity Channel 1 | Eccentricity Channel 2 | Ovality | Glass diameter | **Coating diameter** | Coating eccentricity | Inner coating eccentricity |
| *Nominal* | **260.0** | 0.1 | 0.2 | 3.61 | 125 | **259.9** | 1.3 | 3.8 |
| *Coating -* | **258.4 (-1.6)** | 0.1 | 0.3 | 3.24 | 125 | **257.6 (-2.3)** | 1.7 | 4.3 |
| *Glass +* | **258.0 (-2)** | 0.2 | 0.1 | 11.43 | 130 | **259.1 (-0.8)** | 0.4 | 3.2 |

The differences between coating diameter online test measurements and offline measurements are approximately ±1 µm, but the ovality values obtained from the test measurements are much higher. Particularly in the third measurement, the increase in glass diameter affected the apparent ovality significantly, and the resulting coating diameter appeared still lower than in the first two measurements. The coating to glass eccentricity values are systematically higher in the offline measurements. Additionally, the offline measurements show comparatively high eccentricity of the inner coating layer to the glass layer, however it is not known whether this eccentricity is in the same direction as the outer layer eccentricity.

In the second test, the guiding of the fibre through the coating cup was adjusted such that it induced eccentricity. This was done in approximately the directions of each measurement channel separately. Table 2 shows the result of this test. Eccentricity values are highlighted in bold font, and in the second and third row, the values in brackets show the change compared to the centre position. As in the previous case, ovality is shown as the absolute difference between diameter readings of two measurement channels.

Table 2. Eccentricity variation tests. All values are in µm.

|  | Test measurements | | | | Offline measurements | | | |
|---|---|---|---|---|---|---|---|---|
|  | Coating diameter | **Eccentricity Channel 1** | **Eccentricity Channel 2** | Ovality | Glass diameter | Coating diameter | **Coating eccentricity** | Inner coating eccentricity |
| *Centre* | 260.0 | **0.1** | **0.5** | 3.85 | 125 | 260.1 | **1.5** | 4.3 |
| *Ch1 direction* | 260.3 | **1.3 (+1.2)** | **0.6 (+0.1)** | 6.20 | 125 | 259.9 | **2.7 (+1.2)** | 3.8 |
| *Ch2 direction* | 258.8 | **0.3 (+0.2)** | **1.4 (+0.9)** | 7.44 | 125 | 260.0 | **2.5 (+1)** | 3.6 |

The changes in measured eccentricity values logically followed the induced eccentricity directions, higher values appearing in the corresponding channels. Offline measurements once again showed systematically higher eccentricity values of the outer coatings. However, if the changes from the centre positions are compared (values in bracket in table 1), the difference between online test measurements and offline measurements are within 0.1 µm.

## 5. CONCLUSIONS

The purpose of this work is to study and demonstrate the suitability of the presented method for measurements of geometrical properties of optical fibres during manufacturing. The method is based on the analysis of a scattering pattern, produced when the fibre is illuminated transversely. The method employs two detecting channels, positioned orthogonally. Due to a complicated nature of interference happening between various layers in the fibre, the practical implementation relies on pre-calculated sensitivities of the scattering features and anticipated fibre geometry deviations. For measurements presented in this paper, these calculations were done using an accurate, error-controlled FEM modelling. In principle, simpler or more optimized methods can be employed. We have constructed a prototype instrument and tested it in the production environment with one fibre type.

Two sets of test measurements were performed. In one the effects of diameter variations were studied, and in the other the effects of eccentricity of layers. The experimental results show that the method is capable to detect outer diameter changes below 1 µm, and eccentricity changes in the outer coating layer of 0.1 µm, in online operation. Although the relative changes in geometry follow the offline measurement results, the absolute values for coating diameter and outer coating concentricity deviate from those obtained with offline measurements. In theory, in perfect conditions, and with a fibre without defects, the resolution of the method is much higher still. Sensitivity values obtained from modelling are 60 nm/pixel for the coating diameter and a 15nm/pixel for coating eccentricity.

Eccentricity of the outer coating layer is systematically higher in the offline measurement results. These measurements also show that the inner polymer coating layer is eccentric by several micrometres, and could be affecting the online results. It is unknown whether this eccentricity follows the same direction as the outer layer eccentricity. The effects of the inner coating layer eccentricity on the position of the interference peaks were modelled, and were found to be comparatively small, thus the measurement algorithm does not currently take it into account. However, in this specific case, where the inner coating layer is significantly more eccentric than the outer layer, the effects could be visible, depending on the direction of eccentricity.

The ovality value, calculated as an absolute difference between diameters obtained with two orthogonal measurement channels, is higher than difference to the offline measurement for all tests. One reason for this can be an imperfect information about the distance between the fibre and the detector in one or both channels, caused e.g. by the fibre vibration or improper alignment. It is also possible that this can also be affected by the inner coating layer eccentricity,

The experiments performed on a single fibre type seem promising, and further tests will need to be conducted with various relevant fibre geometries and materials.

*This project 14IND13 PhotInd has received funding from the EMPIR programme co-financed by the Participating States and from the European Union's Horizon 2020 research and innovation programme. The results presented here only reflect authors' view and EURAMET is not responsible for any use that may be made of the information it contains.*